\def\ra{\rangle}
\def\la{\langle}
\def\be{\begin{equation}}
\def\ee{\end{equation}}
\def\ba{\begin{array}}
\def\ea{\end{array}}

\documentstyle[12pt]{article}
\input amssym.def
\begin{document}
\baselineskip=18pt
\setcounter{page}{1}
\centerline{\large\bf Separability of Tripartite Quantum Systems}
\vspace{4ex}
\begin{center}
Ming Li$^{1}$, Shao-Ming Fei$^{1,2}$ and  Zhi-Xi Wang$^{1}$

\vspace{2ex}

\begin{minipage}{5in}

\small $~^{1}$ {\small Department of Mathematics, Capital Normal University, Beijing 100037}

{\small $~^{2}$ Institut f\"ur Angewandte Mathematik, Universit\"at Bonn, D-53115}


\end{minipage}
\end{center}

\begin{center}
\begin{minipage}{5in}
\vspace{1ex}
\centerline{\large Abstract}
\vspace{1ex}

We investigate the separability of arbitrary dimensional tripartite
systems. By introducing a new operator related to
transformations on the subsystems a necessary condition for
the separability of tripartite systems is presented.

\vfill
\smallskip


\end{minipage}\end{center}
\bigskip

Quantum entanglement has been recently recognized as the most
essential ingredient in the quantum information technology. One of the
important problems in the theory of quantum entanglement is the separability.
A multipartite state is called fully separable if and only if the
density matrix $\rho_{AB\cdots C}$ can be written
as: $$\rho_{AB\cdots C}=\sum
\limits_{i}p_{i}\rho_{i}^{A}\otimes\rho_{i}^{B}\otimes\cdots\otimes\rho_{i}^{C},$$
where $\rho_{i}^{A}, \rho_{i}^{B},\cdots,\rho_{i}^{C}$ are density
matrices associated with the subsystems $A,B,\cdots, C$, and $0<p_{i}\leq 1$,
$\sum\limits_{i}p_{i}=1$.

Many separability criteria have been found in recent years. For pure
states, the problem is completely solved, e.g., by using the Schmidt
decomposition $\cite{Werner}$. For mixed states, there are
separability criteria such as PPT, reduction, majorization,
realignment etc. $\cite{Peres,Horodecki,Nielsen,Rudolph,K.Chen}$. In
$\cite{gao}$ the authors have given a lower bound of concurrence for
tripartite quantum states which can be used to detect entanglement.
In $\cite{song}$ the authors have provided a numerically computable
criterion which can detect PPT entangled states for three qubits
systems. The efficient criteron is then generalized to tripartite
systems with arbitrary dimensions $\cite{song2}$. In $\cite{S.J.Wu}$
some nice results shew that some quantity related to Hermitian
matrix is positive for quantum mixed states in $2\times N$ systems,
which was further discussed in $\cite{Rudolph1}$. These results were
generalized to arbitrary dimensional bipartite systems (or $2\times
2\times N$ quantum systems) in $\cite{Zhao}$. In this paper, we
study arbitrary tripartite systems in analogue to the approach used
in $\cite{S.J.Wu,Rudolph1,Zhao}$. The properties of tripartite
density matrices are studied in terms of the Bloch representations.
A necessary condition for the separability of tripartite states has
been obtained. These results are non-trivial when they are reduced
to bipartite systems discussed in $\cite{S.J.Wu,Rudolph1,Zhao}$ and
the separability criterion do detect some entanglements.

Any Hermitian operator on an $N$-dimensional Hilbert space
${\mathcal H}_N$ can be expressed according to the
generators of the special unitary group $SU(N)$ {\cite{Hioe}}.
The generators of $SU(N)$ can be introduced according to the
transition-projection operators
$$P_{jk}=|j\rangle\langle k|,$$
where $|i\rangle$, $i=1,...,N$, are the orthonormal eigenstates of a linear
Hermitian operator on ${\mathcal H}_N$. Set
\begin{eqnarray*}
&&\omega_{l}=-\sqrt{\frac{2}{l(l+1)}}({P}_{11}+{P}_{22}
+\cdots+{P}_{ll}-l {P}_{l+1,l+1}),\\[3mm]
&&{u}_{jk}={P}_{jk}+{P}_{kj},\\[3mm]
&&{v}_{jk}=i({P}_{jk}-{P}_{kj}),
\end{eqnarray*}
where  $1\leq l \leq N-1$ and $1\leq j < k \leq N$. We get a set of
$N^2-1$ operators
$$\Gamma\equiv\{{\omega}_{l},{\omega}_{2},\cdots, {\omega}_{N-1}, {u}_{12},{u}_{13},\cdots, {v}_{12},{v}_{13},
\cdots \},$$
which satisfy the relations
$$Tr\{\lambda_{i}\}=0,\quad\quad
Tr\{\lambda_{i}\lambda_{j}\}=2{\delta}_{ij},~~~\forall~\lambda_i\in\Gamma$$
and thus generate the $SU(N)$ ${\cite{duichenxing}}$.

Any Hermitian operator $\rho$ in ${\mathcal {H}}_{N}$ can be
represented in terms of these generators of $SU(N)$,
\begin{eqnarray}\label{1}
\rho=\frac{1}{N}I_{N}+\frac{1}{2}\sum^{N^{2}-1}_{j=1}r_{j}\lambda_{j},
\end{eqnarray}
where $I_{N}$ is a unit matrix and ${\bf{r}}=(r_{1}, r_{2},\cdots
r_{N^{2}-1})\in {\Bbb{R}}^{N^{2}-1}$. ${\bf{r}}$ is called Bloch
vector. The set of all the Bloch vectors that constitute a
density operator is known as the Bloch vector space
$B({\Bbb{R}}^{N^{2}-1})$.

A matrix of the form (1) is of unit trace and Hermitian, but it
might not be positive. To guarantee the positivity restrictions
must be imposed on the Bloch vector. It is shown
that $B({\Bbb{R}}^{N^{2}-1})$ is a subset of the
ball $D_{R}({\Bbb{R}}^{N^{2}-1})$ of radius
$R=\sqrt{2(1-\frac{1}{N})}$, which is the minimum ball containing
it, and that the ball $D_{r}({\Bbb{R}}^{N^{2}-1})$ of radius
$r=\sqrt{\frac{2}{N(N-1)}}$ is included in $B({\Bbb{R}}^{N^{2}-1})$ {\cite{Harriman}},
that is,
$$D_{r}({\Bbb{R}}^{N^{2}-1})\subseteq B({\Bbb{R}}^{N^{2}-1})\subseteq D_{R}({\Bbb{R}}^{N^{2}-1}).$$

Let the dimensions of systems A, B and C be $N_{1}, N_{2}$ and
 $N_{3}$ respectively. Any tripartite quantum states $\rho_{ABC} \in {\mathcal {H}}_{N_{1}}
  \bigotimes {\mathcal {H}}_{N_{2}} \bigotimes {\mathcal {H}}_{N_{3}}$ can
  be written as:
 \begin{eqnarray}\label{2}
  \rho_{ABC}&=&I_{N_{1}}\otimes I_{N_{2}}\otimes M_{0}+\sum_{i=1}^{N_{1}^{2}-1}\lambda_{i}(1)\otimes I_{N_{2}}\otimes
  M_{i}+\sum_{j=1}^{N_{2}^{2}-1}I_{N_{1}}\otimes \lambda_{j}(2)\otimes \widetilde{M}_{j}\nonumber \\[3mm]
 &&+\sum_{i=1}^{N_{1}^{2}-1}\sum_{j=1}^{N_{2}^{2}-1}\lambda_{i}(1)\otimes \lambda_{j}(2)\otimes
  M_{ij},  \end{eqnarray}
where $\lambda_{i}(1)$, $\lambda_{j}(2)$ are  the generators of
$SU(N_{1})$ and $SU(N_{2})$; $M_{i}, \widetilde{M}_{j}$ and $M_{ij}$
are operators of ${\mathcal{H}}_{N_{3}}$.

{\sf [Theorem 1]} \ Let ${\bf{r}} \in {\Bbb{R}}^{N_{1}^{2}-1}$,
${\bf{s}} \in {\Bbb{R}}^{N_{2}^{2}-1}$ and $|{\bf{r}}| \leq
\sqrt{\frac{2}{N_{1}(N_{1}-1)}}$, $|{\bf{s}}| \leq
\sqrt{\frac{2}{N_{2}(N_{2}-1)}}$. For a tripartite quantum state
$\rho \in {\mathcal {H}}_{N_{1}}
  \bigotimes {\mathcal {H}}_{N_{2}} \bigotimes {\mathcal {H}}_{N_{3}}$
with representation (\ref{2}), we have
\be\label{t1}
M_{0}-\sum_{i=1}^{N_{1}^{2}-1}r_{i}M_{i}-\sum_{j=1}^{N_{2}^{2}-1}s_{j}\widetilde{M}_{j}
+\sum_{i=1}^{N_{1}^{2}-1}\sum_{j=1}^{N_{2}^{2}-1}r_{i}s_{j}M_{ij}\geq
0.
\ee

{\sf[Proof]}\ Since ${\bf{r}} \in {\Bbb{R}}^{N_{1}^{2}-1}$,
${\bf{s}} \in {\Bbb{R}}^{N_{2}^{2}-1}$ and $|{\bf{r}}| \leq
\sqrt{\frac{2}{N_{1}(N_{1}-1)}}$, $|{\bf{s}}| \leq
\sqrt{\frac{2}{N_{2}(N_{2}-1)}}$, we have that $A_{1}\equiv
\frac{1}{2}(\frac{2}{N_{1}}I-\sum\limits_{i=1}^{N_{1}^{2}-1}r_{i}\lambda_{i}(1))$
and $A_{2}\equiv
\frac{1}{2}(\frac{2}{N_{2}}I-\sum\limits_{j=1}^{N_{2}^{2}-1}s_{j}\lambda_{j}(2))$
are positive Hermitian operators. Let
$A=\sqrt{A_{1}}\otimes \sqrt{A_{2}} \otimes I_{N_{3}}$.
Then $A\rho A \geq 0$ and $(A\rho A)^{\dag}=A\rho A$.
The partial trace of $A\rho A$ over ${\mathcal {H}}_{N_{1}}$
(and ${\mathcal {H}}_{N_{2}}$) should be also positive. Hence
\begin{eqnarray*}
0 &\leq& Tr_{AB}(A\rho A) \nonumber \\[3mm]
  &=& Tr_{AB}(A_{1}
\otimes A_{2}\otimes M_{0}+\sum\limits_{i} \sqrt{A_{1}}
\lambda_{i}(1) \sqrt{A_{1}}\otimes A_{2} \otimes
M_{i}+\sum\limits_{j} A_{1} \otimes
\sqrt{A_{2}}\lambda_{j}(2)\sqrt{A_{2}} \otimes \widetilde{M}_{j})\nonumber \\[3mm]
&&+\sum\limits_{ij} \sqrt{A_{1}}\lambda_{i}(1)\sqrt{A_{1}} \otimes
\sqrt{A_{2}}\lambda_{j}(2)\sqrt{A_{2}} \otimes M_{ij}) \nonumber \\[3mm]
  &=&M_{0}-\sum_{i=1}^{N_{1}^{2}-1}r_{i}M_{i}-\sum_{j=1}^{N_{2}^{2}-1}s_{j}\widetilde{M}_{j}
+\sum_{i=1}^{N_{1}^{2}-1}\sum_{j=1}^{N_{2}^{2}-1}r_{i}s_{j}M_{ij}.\nonumber
\end{eqnarray*}
$\hfill\Box$

Formula (\ref{t1}) is valid for any tripartite states. By setting ${\bf s}=0$ in (\ref{t1}),
one can get a result for bipartite systems:

{\sf[Corollary 1]}~ Let $\rho_{AB}\in {\mathcal
{H}}_{N_{1}}\otimes{\mathcal {H}}_{N_{2}}$ which can be generally written as
$\rho_{AB}=I_{N_{1}}\otimes
M_{0}+\sum\limits_{j=1}^{N_{1}^{2}-1}\lambda_{j}\otimes M_{j}$, then
for any ${\bf{r}} \in {\Bbb{R}}^{N_{1}^{2}-1}$ with $|{\bf{r}}| \leq
\sqrt{\frac{2}{N_{1}(N_{1}-1)}}$,
$M_{0}-\sum\limits_{j=1}^{N_{1}^{2}-1}r_{j}M_{j}\geq 0$.

In {\cite{Zhao}}, a separability criterion for $N_{1}\times N_{2}$
systems has been obtained: if $\rho_{AB}$ is separable, then
$M_{0}-\sum\limits_{j=1}^{N_{1}^{2}-1}\displaystyle\frac{4}{3N_{1}-2}d_{j}M_{j}$
is positive for any vector $\overrightarrow{d}=(d_{1}, d_{2},\cdots,
d_{N_{1}^{2}-1})$ with $|\overrightarrow{d}|\leq 1$. Noticing that
$\frac{4}{3N_{1}-2}\leq \sqrt{\frac{2}{N_{1}(N_{1}-1)}}$ for any
$N_{1}\geq 2$, we get from our corollary that this criterion can not
recognize any bipartite entangled states, as it is true for both
entangled and separable states.

A separable state $\rho_{ABC}$ can be written as
$$\rho_{ABC}=\sum\limits_{i} p_{i}| \psi_{i}^{A}\rangle\langle
\psi_{i}^{A}|\otimes| \phi_{i}^{B}\rangle\langle
\phi_{i}^{B}|\otimes| \omega_{i}^{C}\rangle\langle
\omega_{i}^{C}|.$$
From (\ref{1}) it can also be represented as:
\begin{eqnarray}
\rho_{ABC}&=&\sum\limits_{i} p_{i}
\frac{1}{2}(\frac{2}{N_{1}}I_{N_{1}}+\sum\limits_{k=1}^{N_{1}^{2}-1}a_{i}^{(k)}\lambda_{k}(1))
\otimes\frac{1}{2}(\frac{2}{N_{2}}I_{N_{2}}+\sum\limits_{l=1}^{N_{2}^{2}-1}b_{i}^{(l)}\lambda_{l}(2))
\otimes| \omega_{i}^{C}\rangle\langle \omega_{i}^{C}| \nonumber \\
&=&I_{N_{1}}\otimes
I_{N_{2}}\otimes\frac{1}{N_{1}N_{2}}\sum\limits_{i} p_{i}|
\omega_{i}^{C}\rangle\langle \omega_{i}^{C}|
+\sum\limits_{k=1}^{N_{1}^{2}-1}\lambda_{k}(1)\otimes
I_{N_{2}}\otimes\frac{1}{2N_{2}}\sum\limits_{i} a_{i}^{(k)}p_{i}|
\omega_{i}^{C}\rangle\langle \omega_{i}^{C}| \nonumber \\[3mm]
&&+\sum\limits_{l=1}^{N_{2}^{2}-1}I_{N_{1}}\otimes
\lambda_{l}(2)\otimes\frac{1}{2N_{1}}\sum\limits_{i}
b_{i}^{(l)}p_{i}|
\omega_{i}^{C}\rangle\langle \omega_{i}^{C}| \nonumber \\[3mm]
&&+\sum\limits_{k}^{N_{1}^{2}-1}\sum\limits_{l}^{N_{2}^{2}-1}\lambda_{k}(1)\otimes
\lambda_{l}(2)\otimes\frac{1}{4}\sum\limits_{i}
a_{i}^{(k)}b_{i}^{(l)}p_{i}| \omega_{i}^{C}\rangle\langle
\omega_{i}^{C}|,
\end{eqnarray}
where $(a_{i}^{(1)}, a_{i}^{(2)} \cdots, a_{i}^{(N_{1}^{2}-1)})$ and
$(b_{i}^{(1)}, b_{i}^{(2)} \cdots, b_{i}^{(N_{2}^{2}-1)})$ are real
vectors on the Bloch sphere satisfying
$|\overrightarrow{a_{i}}|^{2}=\sum\limits_{j=1}^{N_{1}^{2}-1}(a_{i}^{(j)})^{2}=2(1-\frac{1}{N_{1}})$
and
$|\overrightarrow{b_{i}}|^{2}=\sum\limits_{j=1}^{N_{2}^{2}-1}(b_{i}^{(j)})^{2}=2(1-\frac{1}{N_{2}})$.

Comparing (2) with (4), we have
\begin{eqnarray}
&M_{0}=\frac{1}{N_{1}N_{2}}\sum\limits_{i} p_{i}|
\omega_{i}^{C}\rangle\langle \omega_{i}^{C}|, \quad\quad
M_{k}=\frac{1}{2N_{2}}\sum\limits_{i} a_{i}^{(k)} p_{i}|
\omega_{i}^{C}\rangle\langle \omega_{i}^{C}|, \nonumber \\[3mm]
&\widetilde{M}_{l}=\frac{1}{2N_{1}}\sum\limits_{i} b_{i}^{(l)}
p_{i}| \omega_{i}^{C}\rangle\langle \omega_{i}^{C}|, \quad
M_{kl}=\frac{1}{4}\sum\limits_{i} a_{i}^{(k)}b_{i}^{(l)} p_{i}|
\omega_{i}^{C}\rangle\langle \omega_{i}^{C}|.
\end{eqnarray}

For any $(N_{1}^{2}-1) \times (N_{1}^{2}-1)$ real matrix $R(1)$ and
$(N_{2}^{2}-1)\times (N_{2}^{2}-1)$ real matrix $R(2)$ satisfying
$\frac{1}{(N_{1}-1)^{2}}I-R(1)^{T}R(1)\geq 0$ and
$\frac{1}{(N_{2}-1)^{2}}I-R(2)^{T}R(2)\geq 0$, we define a new
matrix
\begin{eqnarray}\label{5}
\mathcal {R}=\left(%
    \begin{array}{ccc}
      R(1) & 0 & 0 \\
      0 & R(2) & 0 \\
      0 & 0 & T \\
    \end{array}%
    \right),
\end{eqnarray}
where $T$ is a transformation acting on an $(N_{1}^{2}-1) \times
(N_{2}^{2}-1)$ matrix $M$ by $$T(M)=R(1) M R^{T}(2).$$
Using $\mathcal {R}$ we define a new operator $\gamma_{\mathcal {R}}$,
\begin{eqnarray}
  \gamma_{\mathcal {R}}(\rho_{ABC})&=&I_{N_{1}}\otimes I_{N_{2}}\otimes M_{0}^{'}+\sum_{i=1}^{N_{1}^{2}-1}\lambda_{i}(1)
  \otimes I_{N_{2}}\otimes   M_{i}^{'}
  +\sum_{j=1}^{N_{2}^{2}-1}I_{N_{1}}\otimes \lambda_{j}(2)\otimes \widetilde{M}_{j}^{'}\nonumber \\[3mm]
  &&+\sum_{i=1}^{N_{1}^{2}-1}\sum_{j=1}^{N_{2}^{2}-1}\lambda_{i}(1)\otimes \lambda_{j}(2)\otimes
  M_{ij}^{'},
  \end{eqnarray}
where $M_{0}^{'}=M_{0}, \quad
M_{k}^{'}=\sum\limits_{m=1}^{N_{1}^{2}-1}R_{km}(1)M_{m},\quad
\widetilde{M}_{l}^{'}=\sum\limits_{m=1}^{N_{2}^{2}-1}R_{ln}(2)\widetilde{M}_{n}$ and
$M_{ij}^{'}=(T(M))_{ij}=(R(1)MR^{T}(2))_{ij}$.\\

{\sf [Theorem 2]} If $\rho_{ABC}$ is separable, then
$\gamma_{\mathcal {R}}(\rho_{ABC})\geq 0$.\\

{\sf[Proof]} From (5) and (7) we get
\begin{eqnarray*}
M_{0}^{'}&=&M_{0}=\frac{1}{N_{1}N_{2}}\sum\limits_{i} p_{i}|
\omega_{i}^{C}\rangle\langle \omega_{i}^{C}|,~
M_{k}^{'}=\frac{1}{2N_{2}}\sum\limits_{mi} R_{km}(1)a_{i}^{(m)}
p_{i}|\omega_{i}^{C}\rangle\langle \omega_{i}^{C}|,  \\[3mm]
\widetilde{M}_{l}^{'}&=&\frac{1}{2N_{1}}\sum\limits_{ni}
R_{ln}(2)b_{i}^{(n)} p_{i}| \omega_{i}^{C}\rangle\langle
\omega_{i}^{C}|,~ M_{kl}^{'}=\frac{1}{4}\sum\limits_{mni}
R_{km}(1)a_{i}^{(m)}R_{ln}(2)b_{i}^{(n)} p_{i}|
\omega_{i}^{C}\rangle\langle \omega_{i}^{C}|.
\end{eqnarray*}
A straightforward calculation gives rise to
\begin{eqnarray*}\label{3}
  \gamma_{\mathcal {R}}(\rho_{ABC})
  &=&\sum\limits_{i} p_{i}\frac{1}{2}\left(\frac{2}{N_{1}}I_{N_{1}}
             +\sum\limits_{k=1}^{N_{1}^{2}-1}\sum\limits_{m=1}^{N_{1}^{2}-1}R_{km}(1)a_{i}^{(m)}\lambda_{k}(1)\right)\\[3mm]
&&\quad\quad \otimes\frac{1}{2}\left(\frac{2}{N_{2}}I_{N_{2}}
          +\sum\limits_{l=1}^{N_{2}^{2}-1}\sum\limits_{n=1}^{N_{2}^{2}-1}R_{ln}(2)b_{i}^{(n)}\lambda_{l}(2)\right)
\otimes| \omega_{i}^{C}\rangle\langle \omega_{i}^{C}|.
  \end{eqnarray*}
As $\frac{1}{(N_{1}-1)^{2}}I-R(1)^{T}R(1)\geq 0$ and
$\frac{1}{(N_{2}-1)^{2}}I-R(2)^{T}R(2)\geq 0$, we get
$$|\overrightarrow{a_{i}^{'}}|^{2}=|R(1)\overrightarrow{a_{i}}|^{2}\leq\frac{1}{(N_{1}-1)^{2}}|\overrightarrow{a_{i}}|^{2}=
\frac{2}{N_{1}(N_{1}-1)},$$
$$|\overrightarrow{b_{i}^{'}}|^{2}=|R(2)\overrightarrow{b_{i}}|^{2}\leq\frac{1}{(N_{2}-1)^{2}}|\overrightarrow{b_{i}}|^{2}=
\frac{2}{N_{2}(N_{2}-1)}.$$
Therefore $\gamma_{\mathcal {R}}(\rho_{ABC})$ is still a density
operator, i.e. $\gamma_{\mathcal
{R}}(\rho_{ABC})\geq 0$.\hfill$\Box$

Theorem 2 gives a necessary separability criterion for general tripartite
systems. The result can be also applied to bipartite systems.
Let $\rho_{AB}\in{\mathcal {H}}_{N_{1}}\otimes {\mathcal
{H}}_{N_{2}}$, $\rho_{AB}=I_{N_{1}}\otimes
M_{0}+\sum\limits_{j=1}^{N_{1}^{2}-1}\lambda_{j}\otimes M_{j}$. For
any real $(N_{1}^{2}-1)\times (N_{1}^{2}-1)$ matrix $\mathcal {R}$
satisfying $\frac{1}{(N_{1}-1)^{2}}I-{\mathcal {R}}^{T}{\mathcal
{R}}\geq 0$ and any state $\rho_{AB}$, we define $$\gamma_{\mathcal
{R}}(\rho_{AB})=I_{N_{1}}\otimes
M_{0}+\sum\limits_{j=1}^{N_{1}^{2}-1}\lambda_{j}\otimes M_{j}^{'},$$
where $M_{j}^{'}=\sum\limits_{k}{\mathcal {R}}_{jk}M_{k}$.

{\sf [Corollary 2]} For $\rho_{AB}\in{\mathcal {H}}_{N_{1}}\otimes {\mathcal {H}}_{N_{2}}$,
if there exists an ${\mathcal {R}}$ with
$\frac{1}{(N_{1}-1)^{2}}I-{\mathcal {R}}^{T}{\mathcal {R}}\geq 0$
such that $\gamma_{\mathcal {R}}(\rho_{AB})< 0$, then $\rho_{AB}$
must be entangled.

For $2\times N$ systems, our corollary is reduced to the results in \cite{S.J.Wu}.
Generally this criterion do detect certain entanglement of
${\mathcal {H}}_{N_{1}}\otimes {\mathcal{H}}_{N_{2}}$ systems.
As an example we consider the $3\times 3$ Istropic states,
\begin{eqnarray*}
  \rho_{I}&=&\frac{1-p}{9}I_{3}\otimes I_{3}+ \frac{p}{3}\sum\limits_{i,j=1}^{3}|ii\rangle\langle
jj|\\[3mm]
    &=&I_{3}\otimes (\frac{1}{9}I_{3})+\sum\limits_{i=1}^{5}\lambda_{i}\otimes(\frac{p}{6}\lambda_{i})
    -\sum\limits_{i=6}^{8}\lambda_{i}\otimes(\frac{p}{6}\lambda_{i}).
  \end{eqnarray*}
If we choose ${\mathcal {R}}$ to be $\rm{Diag}\{\frac{1}{2},
\frac{1}{2}, \frac{1}{2}, \frac{1}{2}, \frac{1}{2}, -\frac{1}{2},
-\frac{1}{2}, -\frac{1}{2}\}$, we get when $0.5< p \leq 1$,
$\rho_{I}$ is entangled. For tripartite case, we take the following
$3\times3\times3$ mixed state as an example:
\begin{eqnarray*}
\rho=\frac{1-p}{27}I_{27}+p|\psi\ra\la\psi|,
\end{eqnarray*}
where
$|\psi\ra=\frac{1}{\sqrt{3}}(|000\ra+|111\ra+|222\ra)(\la000|+\la111|+\la222|)$.
Taking $R(1)=R(2)=\rm{Diag}\{\frac{1}{2}, \frac{1}{2}, \frac{1}{2},
\frac{1}{2}, \frac{1}{2}, -\frac{1}{2}, -\frac{1}{2},
-\frac{1}{2}\}$, we have that $\rho$ is entangled for
$0.6248<p\leq1$.

In fact the criterion for $2\times N$ systems \cite{S.J.Wu} is equivalent to the PPT
criterion \cite{Rudolph}. Our theorem 2 is also equivalent to the PPT
criterion for $2\times 2\times N$ systems. This can be seen from the followings. Let us choose
   $R(1)=\left(%
    \begin{array}{ccc}
      1 & 0 & 0 \\
      0 & 1 & 0 \\
      0 & 0 & -1 \\
    \end{array}
    \right)$ and $R(2)=I_{3}$.
A separable state $\rho_{ABC}\in {\mathcal {H}}_{2}
  \bigotimes {\mathcal {H}}_{2} \bigotimes {\mathcal
  {H}}_{N}$ can be represented as:
\begin{eqnarray*}
\rho_{ABC}=\sum\limits_{i} p_{i}
\frac{1}{2}(I+\sum\limits_{j=x,y,z}r_{i}^{(j)}\sigma_{j})
\otimes\frac{1}{2}(I+\sum\limits_{j=x,y,z}s_{i}^{(j)}\sigma_{j})
\otimes| \omega_{i}^{C}\rangle\langle \omega_{i}^{C}|.
 \end{eqnarray*}
By the definition we get
\begin{eqnarray*}
  \gamma_{\mathcal {R}}(\rho_{ABC})
  &=&\sum\limits_{i} p_{i}
\frac{1}{2}(I+r_{i}^{(x)}\sigma_{x}-r_{i}^{(y)}\sigma_{y}+r_{i}^{(z)}\sigma_{z})
\otimes\frac{1}{2}(I+\sum\limits_{j=x,y,z}s_{i}^{(j)}\sigma_{j})
\otimes| \omega_{i}^{C}\rangle\langle \omega_{i}^{C}|\nonumber \\
  &=&\rho_{ABC}^{T_{A}}.
  \end{eqnarray*}
For $\rho_{ABC}^{T_{B}}$ and $\rho_{ABC}^{T_{AB}}$, we can similarly choose
$$R(1)=I_{3},~
R(2)=\left(%
    \begin{array}{ccc}
      1 & 0 & 0 \\
      0 & 1 & 0 \\
      0 & 0 & -1 \\
    \end{array}%
    \right)
 \mbox{ and } \quad
R(1)=R(2)=\left(%
    \begin{array}{ccc}
      1 & 0 & 0 \\
      0 & 1 & 0 \\
      0 & 0 & -1 \\
    \end{array}%
    \right). $$
If $\gamma_{\mathcal {R}}(\rho_{ABC})\geq 0$, we get that
$\rho^{T_{A}}, \rho^{T_{B}}$ and $\rho^{T_{AB}}$ are also positive
operators.

On the other hand, if there is an entangled PPT state $\rho_{ABC}\in
{\mathcal {H}}_{2} \bigotimes {\mathcal {H}}_{2} \bigotimes
{\mathcal {H}}_{N}$, $\rho_{ABC}^{T_{A}}\geq 0$,
$\rho_{ABC}^{T_{B}}\geq 0$ and $\rho_{ABC}^{T_{AB}}\geq 0$, but
$\gamma_{\mathcal {R}}(\rho_{ABC})<0$ for some real $3\times 3$
matrices $R(1)$ and $R(2)$ such that $R^{T}(1)R(1)\leq I$ and
$R^{T}(2)R(2)\leq I$, with $\mathcal {R}$ being defined in
(\ref{5}). Then one can define, for all $\vec{\beta}=(\beta_{1},
\beta_{2}, \beta_{3})\in {\Bbb{C}}^{3}$,
$\vec{\beta}^{'}=R(1)\vec{\beta}$ with
$\vec{\beta}^{'}=(\beta_{1}^{'}, \beta_{2}^{'},\beta_{3}^{'})$, and
$\Lambda_{R(1)}(\alpha I+ \sum\limits_{i=1}^{3}\beta_{i}\sigma_{i})
=\alpha I+ \sum\limits_{i=1}^{3}\beta_{i}^{'}\sigma_{i}$. Obviously
$\Lambda_{R(1)}$ would map the Bloch sphere to itself. Hence
$\Lambda_{R(1)}$ is a positive map. From {\cite{S.L. Woronowicz}} it
follows that $\Lambda_{R(1)}$ can be expressed as
$\Lambda_{R(1)}=\Lambda_{R(1)}^{CP}(1)+\Lambda_{R(1)}^{CP}(2)\circ
T$, where $\Lambda_{R(1)}^{CP}(1)$ and $\Lambda_{R(1)}^{CP}(2)$
denote completely positive maps, and $T$ the transpose. Similar
result can be obtained for $\Lambda_{R(2)}$. A straightforward
calculation shows that
\begin{eqnarray*}
  \gamma_{\mathcal {R}}(\rho_{ABC})
  &=&(\Lambda_{R(1)}\otimes \Lambda_{R(2)} \otimes I)(\rho_{ABC}) \nonumber
  \\[3mm]
  &=&((\Lambda_{R(1)}^{CP}(1)+\Lambda_{R(1)}^{CP}(2)\circ
T)\otimes (\Lambda_{R(2)}^{CP}(1)+\Lambda_{R(2)}^{CP}(2)\circ T)
\otimes I)(\rho_{ABC})\nonumber \\[3mm]
  &=&(\Lambda_{R(1)}^{CP}(1) \otimes \Lambda_{R(2)}^{CP}(1) \otimes
  I)(\rho_{ABC})+(\Lambda_{R(1)}^{CP}(2) \otimes \Lambda_{R(2)}^{CP}(1) \otimes
  I)(\rho_{ABC}^{T_{A}})\nonumber \\[3mm]
  &&+(\Lambda_{R(1)}^{CP}(1) \otimes \Lambda_{R(2)}^{CP}(2) \otimes
  I)(\rho_{ABC}^{T_{B}})
  +(\Lambda_{R(1)}^{CP}(2) \otimes \Lambda_{R(2)}^{CP}(2) \otimes
  I)(\rho_{ABC}^{T_{AB}}).
  \end{eqnarray*}
Since that the tensor product of two completely positive maps is
still a completely positive map and $\rho_{ABC}^{T_{A}}\geq 0$,
$\rho_{ABC}^{T_{B}}\geq 0$ and $\rho_{ABC}^{T_{AB}}\geq 0$, this
implies that $\gamma_{\mathcal {R}}(\rho_{ABC})\geq 0$. This is a
contradiction. Hence the theorem 2 is equivalent to the PPT criterion
for $2\times 2\times N$ systems.

\bigskip
We have studied the separability of tripartite quantum systems. In
terms of the Bloch representation of density matrices, a necessary
condition for the separability of tripartite states has been
obtained. Our approach gives a new way of separability
investigation. For $2\times2\times N$ systems our criterion is
equivalent to PPT, namely PPT criterion can be also understood
according to the Bloch representation approach. Nevertheless it is
rather complicated to compare our criterion with PPT generally for
higher dimensional tripartite systems. Moreover as the PPT and
realignment separability criteria can give rise to lower bonds of
entanglement of formation and concurrence \cite{Chen-Albeverio-Fei},
one could also discuss the possible relations between the lower
bounds of entanglement and the separability criterion in this
letter. The approach can be also generalized to arbitrary
multipartite systems.

\bigskip
\noindent{\bf Acknowledgments}\, The work is partly supported by
NKBRPC(2004CB318000).

\end{document}